# *naplib-python*: Neural Acoustic Data Processing and Analysis Tools in Python

Gavin Mischler [1, 2]
  gm2944@columbia.edu

Vinay Raghavan [1, 2]
  vsr2119@columbia.edu

Menoua Keshishian [1, 2]
  mk4011@columbia.edu

Nima Mesgarani [1, 2, 3]
  nima@ee.columbia.edu

[1] Mortimer B. Zuckerman Mind Brain Behavior, Columbia University, New York, United States
[2] Department of Electrical Engineering, Columbia University, New York, United States
[3] Corresponding author

## Abstract
Recently, the computational neuroscience community has pushed for more transparent and reproducible methods across the field. In the interest of unifying the domain of auditory neuroscience, naplib-python provides an intuitive and general data structure for handling all neural recordings and stimuli, as well as extensive preprocessing, feature extraction, and analysis tools which operate on that data structure. The package removes many of the complications associated with this domain, such as varying trial durations and multi-modal stimuli, and provides a general-purpose analysis framework that interfaces easily with existing toolboxes used in the field.

## Keywords
python, auditory neuroscience, iEEG, ECoG, preprocessing

## Code metadata

| Nr | Code metadata description | Please fill in this column |
|---|---|---|
| C1 | Current code version | 0.2.0 |
| C2 | Permanent link to code/repository used for this code version | *https://github.com/naplab/naplib-python* |
| C3 | Permanent link to reproducible capsule | *https://codeocean.com/capsule/6656601/tree* |
| C4 | Legal code license | MIT License |
| C5 | Code versioning system used | git |
| C6 | Software code languages, tools and services used | python |
| C7 | Compilation requirements, operating environments and dependencies | Linux, macOS, or Windows; matplotlib, numpy, scipy, pandas, statsmodels, hdf5storage, mne, scikit-learn |
| C8 | If available, link to developer documentation/manual | *https://naplib-python.readthedocs.io* |
| C9 | Support email for questions | *nima@ee.columbia.edu* |

# 1. Introduction

With the recent explosion of neural data acquisition and computational power, the field of neuroscience has seen incredible growth in the use of computational methods to analyze neural response patterns and make inferences about the brain. The field of auditory neuroscience is no exception, with the widespread use of computational analyses such as spectro-temporal receptive field (STRF) estimation (Aertsen et al., 1981; Theunissen et al., 2000, 2001) and software such as STRFlab (http://www.strflab.berkeley.edu), mTRF (Crosse et al., 2016), Neural Encoding Model System (NEMS) (David, 2018), and others dedicated to these specific techniques. Many papers are now accompanied by small bits of code to reproduce figures or analyses, which greatly aids in the reproducibility of scientific research. However, the explosion of computational methods has also led to a software ecosystem with many highly specialized packages written by, and for, people with very different needs. Even when code is shared openly, it may do little to aid the reproducibility of the experiments because of difficulties running others' code or omissions of critical details in the original report (Easterbrook, 2014; Miłkowski et al., 2018). Therefore, there is a need for a unifying framework which provides comprehensive tools for the auditory neuroscience domain and can easily fit into existing codebases and analysis pipelines used by researchers in the field.

As a **N**eural **A**coustic **P**rocessing **Lib**rary in Python, naplib-python is specifically built for auditory neuroscience. Its library of methods complements those provided by other public toolkits. Python is an ideal language for this purpose, due to its vast open-source ecosystem and the fact that many in the neuroscience community are moving towards Python (Muller et al., 2015). In addition to implementing a host of relevant analysis methods that previously had no popular Python implementation, naplib-python expands upon the existing MATLAB toolkit NAPlib (Khalighinejad et al., 2017), which focuses on phoneme response analysis methods, by porting several of these methods to Python and increasing their ease-of-use by ensuring they work with naplib-python data structures. One existing Python package, MNE-Python (Gramfort et al., 2013), supports a broad set of functionalities for neural data, including magnetoencephalography (MEG), electroencephalography (EEG), and intracranial EEG (iEEG). However, that package was never meant to be used only for auditory neuroscience, and so it generally offers techniques applicable to many areas of neuroscience while missing functionality that would be useful in the auditory domain, such as linguistic alignment and phonetic feature extraction. In this paper, we first describe the basic data structure and API used in the package, then we provide an overview of the methods available in the package and describe their use in common analysis pipelines in the field.

## 2. Package Description

### *2.1. Data Structure and API*

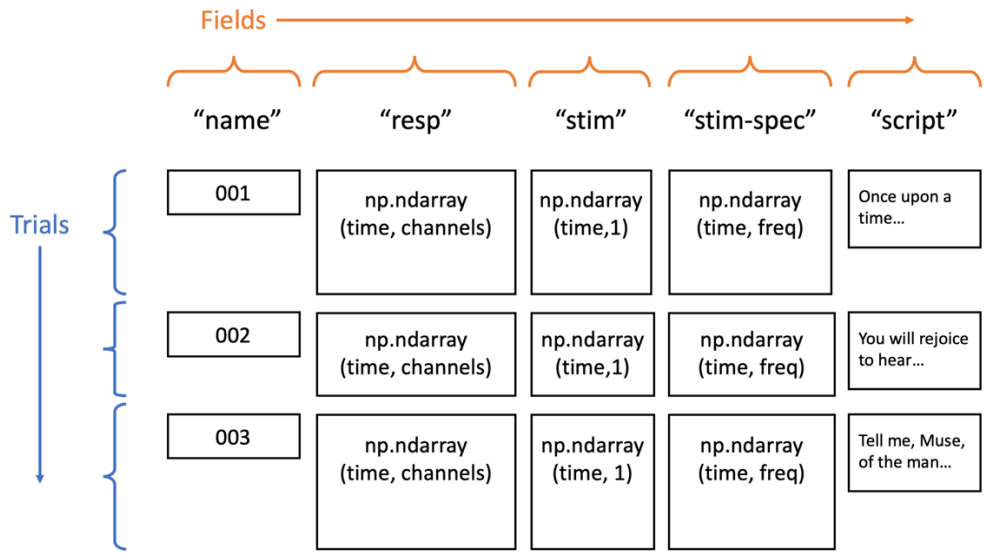

Fig. 1. Basic Data object structure for an example task where neural responses to natural language stories were recorded. None of the fields depicted is required for the Data object, but they naturally illustrate what might be included in a typical task.

In auditory neuroscience, several problems often arise which make using a general-purpose data structure difficult. Stimuli tend to be non-uniform in duration, especially when using naturalistic stimuli like human speech (Hamilton & Huth, 2020). Additionally, a large variety of trial-specific metadata may be needed for analyses, such as transcripts of speech stimuli or event labels, which would traditionally need to be loaded and treated as separate variables. The Data object in naplib-python takes care of these problems by seamlessly storing any number of trials containing any number of fields, which may include things like auditory stimulus waveforms and time-frequency representations, neural responses, stimulus transcripts, and trial condition labels. By not enforcing equal durations and allowing various data types through its various fields, the Data class is reminiscent of a pandas DataFrame (McKinney, 2011), and in many ways it operates similarly. Fig. 1 shows a visual representation of the most typical information that could be stored in a Data instance. The Data object is designed to enable easy processing by trial, by field, or all together. Many functions in naplib-python can be called by either passing a set of parameters, or simply passing a Data object containing all the necessary fields to fill the parameters. For example, a user can fit a STRF model using the TRF class in naplib-python by passing in a Data instance and the fields for stimulus and response will be automatically extracted and used as the input and output when training the model. Alternatively, certain parameters can be passed in manually without needing them to be stored in a Data instance.

For convenient adoption, naplib-python supports loading data from the Brain Imaging Data Structure (BIDS) format (Gorgolewski et al., 2016; Niso et al., 2018; Pernet et al., 2019) directly into a Data object with a single line of code. Trials are automatically separated based on events defined in the BIDS files, and then the data is ready to be quickly processed. These features make the Data object easy for beginners to adopt and broadly useful for any type of analysis.

## 2.2. Library Overview

In this section we summarize the main modules currently available in the package and the tools offered in each. Full documentation and examples for the functions within these modules, as well as the other utility modules, are available online at https://naplib-python.readthedocs.io.

### 2.2.1. Features

In auditory neuroscience, a wide variety of features have been proposed to describe both acoustic signals and neural data. An implementation of the auditory spectrogram (Yang et al., 1992) is provided, a time-frequency representation which models the inner ear and cochlear spectral decomposition of sound waves. Additionally, many linguistic features are available to describe speech signals. A forced aligner is provided based on the Prosodylab-Aligner (Gorman et al., 2011), as well as functions to extract phoneme and word alignment labels from the aligner's output, which can be used to identify phonetic information and timing in speech stimuli.

### 2.2.2. Encoding

This module is dedicated to encoding models used by the auditory neuroscience community. For example, a robust temporal receptive field (TRF) class is implemented which interfaces naturally with the Data object. By default, the TRF model uses cross-validated ridge regression, but any class which adheres to the scikit-learn linear model API (Pedregosa et al., 2011) can be used, meaning that TRFs can be trained from L1-regularized models, elastic net models, or any other linear or non-linear user-defined classes. This is useful for training both forward STRF models (Theunissen et al., 2000, 2001), which predict neural responses from acoustic stimuli, or backward models for stimulus reconstruction (Bialek et al., 1991; Mesgarani et al., 2009), which reconstruct acoustic stimuli from neural signals.

### 2.2.3. Segmentation

Stimulus onset response patterns are often studied in order to understand response properties in auditory neuroscience, as used in studies of evoked potentials (Gage et al., 1998; Näätänen et al., 1978; Picton, 2013) or stimulus onset-locked encoding (Gwilliams et al., 2018; Phillips et al., 2002; Hamilton et al., 2018). The segmentation module contains methods for segmenting multi-trial data based on aligned labels. For example, aligned labels could include phoneme onset labels (where phoneme alignment can be computed using the Features module described above), enabling easy analysis of phoneme onset responses.

### 2.2.4. Preprocessing

A significant amount of preprocessing is typically involved when analyzing neural signals, so this module includes several functions which are useful for a variety of signal types, especially intracranial recordings. There are functions for extracting the envelope and phase of different frequency bands using a filter bank followed by the Hilbert transform (Edwards et al., 2009), which can be used to extract the well-studied high-gamma envelope response for iEEG data, or as an input to further analyses of phase-amplitude coupling (Canolty et al., 2006; Tort et al.,

2010). There are also generic filtering functions that operate on Data instances which are useful for performing notch filtering to remove line noise or filtering EEG/MEG data into different frequency bands.

### 2.2.5. Stats

Statistical analysis of data is critical to understand the significance of any findings in neuroscience. This module provides several statistical tools common to auditory neuroscience. For example, one of the first steps in many analysis pipelines is electrode selection, which can be done using a t-test between responses to speech and silence to identify stimulus-responsive electrodes (Mesgarani & Chang, 2012). Another common statistic offered in the package and used in the field is the F-ratio, which is often used to describe the discriminability of neural responses between different stimulus classes (Khalighinejad et al., 2021). Additionally, a linear mixed effects model is offered to perform linear modeling with the ability to control for effects such as subject identity, which may be needed when studying data across heterogeneous subjects, as is common with iEEG data. Similarly, a generalized t-test method is offered, which can be used to perform t-tests while controlling for additional factors, such as subject identity when testing a distribution with underlying groupings.

### 2.2.6. Input-Output

Input from and output to files is supported in the IO module, including functions to save and load files, as well as read from or write to third-party file structures like MATLAB and BIDS format. These functions make naplib-python easy to use no matter how a researcher's data is currently stored or what stage of the analysis pipeline a researcher wants to use naplib-python.

## 3. Software Impact

With its large suite of implemented methods, naplib-python enables researchers in the field of auditory neuroscience to run common analysis pipelines in only a few lines of code, all while using a general data framework that applies to nearly any type of neural recording data. This allows for collaborations and easier code sharing across disciplines, even when data or recording methods differ significantly between researchers. Furthermore, the ability to interface between naplib-python and other commonly used packages greatly extends the utility of all related toolkits, since researchers can rely on the naplib-python framework and data structure for basic analyses but still utilize state-of-the-art methods available elsewhere without needing to write new code. There are multiple tutorial notebooks available on the online documentation illustrating how to integrate naplib-python with other toolkits, such as fitting encoding models on naplib-python data using NEMS and plotting EEG analysis results using MNE visualization tools.

In addition to offering an analysis framework which can interface with various third-party packages, naplib-python offers a wide array of support for methods which are commonly used in the field but lack standard open-source implementations. This means naplib-python is well-positioned to become the standard toolbox for researchers in the field. Introducing this package to the field will therefore improve the reproducibility of new research by reducing the amount of independently-produced code and making code-sharing easier. The package was critical in a recent study of noise adaptation mechanisms in auditory cortex (Mischler et al., 2022).

## 4. Limitations and Future Improvements

The main limitation of naplib-python is because all operations are performed in-memory, processing of large (hours-long) datasets becomes highly inefficient. This could be enhanced in the future with dynamic data loading and saving for individual trial data. Additionally, while data can be imported from various sources, there is currently limited ability to save data to a wide variety of file structures, since naplib-python is primarily useful for the later stages of data analysis, beginning with preprocessing after raw data have been collected and stored.

**Acknowledgements**
*We thank the members of the Neural Acoustic Processing Lab who provided feedback on the package and its applications. This work was supported by National Institutes of Health grant R01DC018805 and National Institute on Deafness and Other Communication Disorders grant R01DC014279. GM was supported in part by the National Science Foundation Graduate Research Fellowship Program under grant DGE2036197.*